\documentclass[aps,prl,twocolumn,floats,showpacs,superscriptaddress]{revtex4-1}
\usepackage{graphicx,epsfig,psfrag}
\usepackage{times}
\usepackage{graphics,dcolumn,bm,float}
\usepackage{amssymb,amsmath,rotate,color}

\usepackage{graphicx}

\begin{document}
\unitlength 1 cm
\newcommand{\be}{\begin{equation}}
\newcommand{\ee}{\end{equation}}
\newcommand{\bearr}{\begin{eqnarray}}
\newcommand{\eearr}{\end{eqnarray}}
\newcommand{\nn}{\nonumber}
\newcommand{\la}{\langle}
\newcommand{\ra}{\rangle}
\newcommand{\cd}{c^\dagger}
\newcommand{\vd}{v^\dagger}
\newcommand{\ad}{a^\dagger}
\newcommand{\bd}{b^\dagger}
\newcommand{\tk}{{\tilde{k}}}
\newcommand{\tp}{{\tilde{p}}}
\newcommand{\tq}{{\tilde{q}}}
\newcommand{\eps}{\varepsilon}
\newcommand{\vk}{{\vec k}}
\newcommand{\vp}{{\vec p}}
\newcommand{\vq}{{\vec q}}
\newcommand{\vkp}{\vec {k'}}
\newcommand{\vpp}{\vec {p'}}
\newcommand{\vqp}{\vec {q'}}
\newcommand{\bk}{{\bf k}}
\newcommand{\bp}{{\bf p}}
\newcommand{\bq}{{\bf q}}
\newcommand{\br}{{\bf r}}
\newcommand{\bR}{{\bf R}}
\newcommand{\up}{\uparrow}
\newcommand{\down}{\downarrow}
\newcommand{\fns}{\footnotesize}
\newcommand{\ns}{\normalsize}
\newcommand{\cdag}{c^{\dagger}}
\newcommand{\lc}{\langle\!\langle}
\newcommand{\rc}{\rangle\!\rangle}

\title{Kondo resonance from vacancies in graphene}
\author{S. A. Jafari}
\affiliation{Department of Physics, Sharif University of Technology, Tehran 11155-9161, Iran}
\affiliation{Center of excellence for Complex Systems and Condensed Matter (CSCM), Sharif University of Technology, Tehran 1458889694, Iran}
\affiliation{School of Physics, Institute for Research in Fundamental Sciences (IPM), Tehran 19395-5531, Iran}
\author{T. Tohyama}
\affiliation{Yukawa Institute for Theoretical Physics, Kyoto University, Kyoto 606-8502, Japan}

\begin{abstract}
Using the slave-rotor mean-field theory, we study the formation of Kondo resonance
in graphene induced from vacancies. At the Dirac neutrality point, we find a
critical hybridization strength beyond which the Kondo resonance takes place,
despite vanishing density of states. 
We also find that the line-shapes of the Kondo peak are entirely
different from Lorentzian shown in normal metallic hosts.
The dependence of Kondo temperature on the
effective hybridization parameter of the spinon field turns out
to be linear in contrast to the quadratic dependence in normal
metals. Upon doping we find strong electron-hole asymmetry for
the dependence of Kondo temperature on the chemical potential.
\end{abstract}

\maketitle

{\em Introduction}: 
Graphene and its underlying effective Dirac theory already
at its single-particle level serves as a source of many exciting 
phenomena in condensed matter physics~\cite{Geim} which do not have a counterpart 
in ordinary metals as they are driven from a relativistic structure
of the theory, albeit for 1 eV energy scales~\cite{NetoRMP}. 
Even some phenomena such as quantum Hall states
which are common in graphene and two-dimensional electron gas enjoy
the benefit of being realizable at ordinary experimental conditions
when it comes to graphene~\cite{Kim}. The dangling $2p_z$ orbitals in graphene
provide a suitable platform to host external ad-atoms~\cite{Danny}. Its two dimensional
crystal structure facilitates the creation of vacancies which are
believed to be source of spin-half magnetic moments~\cite{Nair}. 
Localized magnetic states in solids can form if a discrete energy level
with considerably large Coulomb repulsion, $U$, hybridizes with the 
continuum of states in the solids~\cite{AndersonSIAM}. 
Once such magnetic doublets
(corresponding to $\up$ and $\down$ orientations) are formed, below a
certain energy scale $k_BT_K$ tunneling between the two  
screens out the magnetic moment, leading to the Kondo effect~\cite{Hewson}.

When the host solid
is a simple metal, the resulting Kondo problem has been much studied
within both single-impurity Anderson model and its low-energy effective
theory of the Kondo model. 
However when the host is graphene~\cite{Novoselov1,Novoselov2}, certain features 
of graphene can lead to quite distinct Kondo physics. First of all, 
unlike normal metals the dispersion relation characterizing the continuum of 
states in graphene is given by a cone-shaped relation $\eps_\vk=\hbar v_F|\vk|$. 
This gives rise to a V-shaped density of states (DOS), linearly vanishing
at the Dirac neutrality point at the center of the $\pi$-band. 
When the chemical potential is tuned to the right (left)
side of the Dirac point, the system is electron (hole) doped.
The Dirac point itself is singular in the sense that it is neither
electron- nor hole-doped. 
In addition, the two-dimensional structure of graphene brings about a new 
type of localized states which, instead of coming from $d$ or $f$ orbital of 
an external transition metal, can arise from its own vacancies~\cite{chen}. 
At the Hartree mean field level, localized magnetic states form in graphene 
when a discrete level is coupled to its Dirac fermions~\cite{Uchoa}.
When the localized states arise from a vacancy or an ad-atom in the hollow-site, 
the magnetic states become even robust~\cite{Mashkoori,Uchoa2}.
The study of Kondo model shows that pseudo-gap like structure of 
the DOS of graphene around the Dirac point leads to 
asymmetric scaling of the Kondo temperature with the 
doping $\mu$ as $T_K\propto |\mu|^{x_\pm}$, where two 
different exponents $x_\pm$ characterize the vanishing of Kondo 
temperature as a function of doping level $\mu$ depending on 
whether it is approached from electron-doped side or the hole-doped 
side~\cite{Bulla}. In this work within the Anderson model we will
find a totally different behavior for small $\mu$, and that moreover
the Dirac point ($\mu=0$) itself allows for non-zero Kondo temperature.

The simplest way to understand the formation of Kondo singlet
in metals is in terms of a variational wave-function where the binding
energy of the singlet gives the Kondo temperature as
$T_K\propto \exp\left[-1/(J\rho_0) \right]$, 
where $J$ is the exchange coupling between the conduction electrons
and the localized magnetic moment formed at the impurity site and
$\rho_0$ is the DOS of the host metal at the Fermi level~\cite{Yosida}.
This simple formula for the normal metals implies that, when 
the DOS $\rho_0$ becomes zero, the Kondo temperature $T_K$ is also
expected to vanish. Some of the existing theoretical 
investigations~\cite{Bulla,Berakdar,Balseiro} support this view.
However the recent experimental data on vacancy induced Kondo
effect in graphene indicates that not only the Kondo temperature
at the Dirac point is non-zero, but the typical scale of
Kondo temperatures is larger than those in typical metals~\cite{chen}.
Previous study of the single impurity Anderson model using numerical renormalization group 
method predicts the possibility of a non-zero Kondo temperature when the 
hybridization strength $V$ is larger than a critical value on the scale 
of $7$ eV, which does not seem to be feasible in graphene. 

In this letter, we study the single impurity Anderson model by means of 
slave rotor theory~\cite{DSR02,DSR04}. We assume that the 
impurity level arises from a Jahn-Teller distorted vacancy and 
the hybridization has $s-wave$ symmetry.  Within the mean field slave rotor
theory we find that the Kondo resonance can form even at the Dirac
neutrality point under conceivable conditions for the critical hybridization
strength. 
We also find that, when the localized state is due to a vacancy in graphene
lattice, the resonance line-shape is entirely different from the 
conventional Lorentzian shape that gives rise to a linear dependence
of the Kondo temperature on the effective hybridization in contrast to the
quadratic dependence of ordinary metals. Away from the Dirac neutrality point, 
we find analytic expressions for the chemical potential dependence of the Kondo 
temperature that exhibits strong particle-hole asymmetry.

{\em Formulation of the problem:}
Florens and Goerges presented a slave rotor formulation 
of the Anderson impurity problem~\cite{DSR02,DSR04} that is appropriate 
when the Hubbard term is in general SU(N) symmetric,
\be
   H_{\rm local}= \sum_\sigma (\eps-\mu) d^\dagger_\sigma d_\sigma
   +\frac{U}{2}\left[\sum_\sigma d^\dagger_\sigma d_\sigma-1\right]^2.
\ee
In our case $\sigma$ denotes the SU(2) spin, $d^\dagger_\sigma$ creates
a physical electron in the impurity level and $\eps=\eps_d+U/2$.
Decomposing the electron creation operator as
$d^\dagger_\sigma=f^\dagger_\sigma e^{i\theta}$, where $f_\sigma$ is a spinon
carrying the spin and $\theta$ is a rotor field conjugate to the occupancy
of the localized site, 
enlarges the Hilbert space. The physical Hilbert space corresponds
to parts of the enlarged Hilbert space, where the pseudo fermion field 
(spinon) and the angular momentum of the rotor field  $\hat L$ are locked by the constraint 
$\hat L =\sum_\sigma\left[f^\dagger_\sigma f_\sigma-\frac{1}{2} \right]$.
The rotor field $\theta$ is conjugate to charge variable, and
the angular momentum $\hat L$ generates shifts in $\theta$. 
Upon the mean field decoupling of the rotor ($\theta$) and
spinon ($f$) fields, one ends up with the following two simple
effective Hamiltonians in the spinon and rotor sectors~\cite{DSR04}:
\bearr
   &&H_f=(\xi-h)\sum_\sigma f^\dagger_\sigma f_\sigma
   +\sum_{\vk\sigma} \Phi_\vk a^\dagger_{\vk\sigma}b_{\vk\sigma}+ h.c.
   \label{Hf.eqn}
   \\
   &&+\sum_{\vk\sigma}\tilde V_{\vk} a^\dagger_{\vk\sigma}f_\sigma+
   \tilde V_\vk^* f^\dagger_\sigma a_{\vk\sigma}-
   \mu\sum_{\vk} \left( a^\dagger_{\vk\sigma}a_{\vk\sigma}
   +b^\dagger_{\vk\sigma}b_{\vk\sigma}\right) \nn
\eearr
and
\be
   H_\theta=\frac{U}{2}\hat L^2+h\hat L + K\cos\theta
   \label{Htheta.eqn},
\ee
where $\xi=\eps-\mu$ and $h$ is the Lagrange 
multiplier that imposes the constraint. 
$a^\dagger_{\vk\sigma}$ and $b^\dagger_{\vk\sigma}$ are creation
operators corresponding to A and B sublattices of the honeycomb lattice, respectively. 
$\Phi_\vk$ is the standard form factor for the honeycomb lattice,
which in a low-energy theory becomes $\Phi_\vk=v_F(k_x+ik_y)$,
where $v_F$ is the Fermi velocity in graphene.
Note that here we have assumed that the hybridization takes place only 
between the local orbital and the A-sublattice fermions.
The parameters $\tilde V$ and $K$ of the decoupled Hamiltonians 
satisfy the following equations:
\be
   \tilde V_\vk = V_\vk\langle \cos\theta\rangle_\theta,~~~~~
   K=\sum_{\vk\sigma}  V^*_\vk\langle a_{\vk\sigma}f^\dagger_\sigma  \rangle_f,
\ee
along with the constraint
$ \langle \hat L\rangle_\theta=2\left[n_F(\xi-h)-\frac{1}{2}\right]$, 
the graphical solution of which at very low-temperatures
indicates that the Lagrange multiplier $h=\xi$. 

As noticed in Ref.~\cite{DSR04}, in the small $V$ regime 
and for metallic hosts, the perturbative treatment within the 
slave rotor formulation gives the exact dependence of the width of 
the Kondo resonance on the parameters $U$ and $V$ of the original 
Anderson model~\cite{DSR04}. To specialize the Anderson problem
for the case of vacancy in graphene, we build on the proposal in
Ref.~\cite{ogata}, according to which the localized orbital
in graphene vacancies responsible for the Kondo effect is formed
by a combination of neighboring $\sigma$ orbitals. Then
the hybridization $V$ is expected to be small, and hence the
perturbative approach of Ref.~\cite{DSR04} is expected to work well.
Having assumed that the impurity is due to a carbon vacancy,
the resulting localized state is expected to be hybridized 
with three neighboring atoms. Therefore the s-wave symmetric
hybridization function is given by~\cite{Uchoa2}
\be
   \tilde V_{\vk}=\tilde V\frac{\Phi_\vk}{v_F}.
\ee
Such a $\vk$ dependence turns out to be crucial since
it gives rise to anomalous hybridization function as well as
normalizable spectral function.

In the rest of the paper we proceed by finding a self-consistent
solution for the coupled spinon-rotor Hamiltonians. To find the 
ground state of the rotor sector, we note that at small $V$ we expect $K$  
to be small and hence a perturbation theory around $K=0$ point can be applicable. 
The unperturbed spectrum is given by $E^{(0)}_\ell=(\frac{U}{2}\ell+\xi)\ell$,
and the corresponding eigenstate is given by 
$\langle \theta|\ell\rangle=e^{i\ell \theta}/\sqrt{2\pi}$.
Assuming that the condition $-U<\eps_d-\mu<0$ or 
equivalently $2|\xi|<U$ holds, the ground state
of this spectrum corresponds to $\ell=0$ and hence $E_0^{(0)}=0$.
Note that, by virtue of the constraint locking the spinon and
rotor parts, the $|\ell=0\rangle$ ground state for the rotor sector
is equivalent to single-occupancy of the impurity state that is a
pre-requisite for the formation of Kondo resonance. The perturbation
$K\cos\theta$ modifies the rotor ground state in the first order
such that the expectation value of $\cos\theta$ in the
first-order corrected ground state becomes,
$
   \langle \cos\theta\rangle_\theta=\frac{2UK}{4\xi^2-U^2}.
$.
Being combined with the one defining $K$, this equation gives the 
following self-consistency condition~\cite{DSR04}:
\be
   \frac{\tilde V^2}{V^2}\frac{4\xi^2-U^2}{2U}=
   \sum_{\vk\sigma}\tilde V^*_\vk\langle a_{\vk\sigma}f^\dagger_\sigma  \rangle_f
   \label{affa.eqn}.
\ee
To calculate the expectation value required in the right hand side (RHS)of the above
equation, we write down the equation of motion for the resonance 
Hamiltonian~\eqref{Hf.eqn} of the spinons, which gives
\be
   \lc a_{\vk\sigma}|f^\dagger_\sigma\rc=
   \frac{\tilde V}{v_F}\frac{z\Phi_\vk}{z^2-|\Phi_\vk|^2}\frac{1}{z-\mu-\Delta(z)}
\ee
where $z=\omega+\mu$, and the hybridization function becomes
\bearr
   &&\Delta(z)=\frac{\tilde V^2}{v_F^2}\sum_{\vp}\frac{z|\Phi_\vp|^2}{z^2-|\Phi_\vp|^2}=
   \label{anomal.eqn}\\
   &&-\frac{\pi|\tilde V|^2z}{D^4} \left [
   D^2+z^2\ln\frac{D^2-z^2}{z^2} \right ] 
   -i\frac{\pi^2|\tilde V|^2z^2|z|}{D^4}\nn.
\eearr
The momentum cutoff $\Lambda=\sqrt{2\pi}$ corresponding to one-half 
electron per allowed momentum gives the energy cutoff
$D=v_F\Lambda=v_F\sqrt{2\pi}$. 
Note that, when the impurity atom adopts a "top-site" position and
hybridizes with only one carbon atom beneath itself, the hybridization 
function becomes wave-vector independent and the imaginary part of $\Delta(z)$
will be proportional to $|z|$, i.e., the DOS, instead of the above anomalous form $z^2|z|$ that is the case for vacancies. To see the consequences of this form, 
let us express the self-consistency condition~\eqref{affa.eqn}
in terms of local spinon spectra,which can be done by using the 
spectral theorem,
\bearr
   &&\rho_{f}(z)=-\frac{1}{\pi}
   \frac{\Delta_I(z)}{[z-\mu-\Delta_R(z)]^2+\Delta^2_I(z)},
   \label{rhof.eqn}\\
   &&\frac{\tilde V^2}{V^2}\frac{4\xi^2-U^2}{2U}=
   \int_{-\infty}^\mu dz (z-\mu) \rho_f(z)
   \label{self3.eqn},
\eearr
where $\rho_f$ is the spectral density of spinons.

When the parameters $V$ is not large -- which holds for
graphene with vacancies --, for moderate to large values of $U$
the ground state of the rotor sector is expected to be
slowly varying function of $\theta$ as the charge fluctuations
are not strong. Hence the rotor
correlator is expected to be substantial only for zero
frequency, which means
$\lc e^{i\theta(\tau)}e^{-i\theta(0)}\rc_\theta \approx 1$.
Therefore the local Green's function of physical electrons is 
essentially given by that of the spinons: $g_d(\tau)=g_f(\tau)$.
Within this formulation, the line-shape of the Kondo
resonance is approximately given by~\eqref{rhof.eqn}.
For typical metals the energy dependence of the imaginary part 
of the hybridization function, $\Delta_I$, follows the energy dependence
of the DOS of the host and therefore is almost a constant such that 
the Kondo line-shape for metallic hosts can be fitted
by a Lorentzian function. In the case of vacancy, local orbital 
hybridizing with Dirac spectrum of graphene, $\Delta_I$, has
an anomalous strong energy dependence given by~\eqref{anomal.eqn}.
Hence within the slave rotor mean field the following picture 
for the line-shape of the Kondo resonance in graphene emerges: 
The line-shape is essentially derived from~\eqref{rhof.eqn} and 
is controlled by the parameter $\tilde V$. This parameter in turn
is related to the bare parameters of the original Anderson Hamiltonian via 
self-consistency~\eqref{self3.eqn}. To proceed further, let us
normalize all energy parameters in units of the bandwidth $D$ and
write the above equations as,
\bearr
   \rho_{f}(z) &=&\frac{1}{\pi}\frac{cz^2|z|}{(z-b)^2+c^2z^6}\\
   \frac{4\zeta^2-u^2}{4 v^2 u}&=&\frac{1}{\pi\tilde v^2D} \int_{-D}^\mu 
   \frac{cz^2|z|(z-\mu)dz}{(z-b)^2+c^2 z^6} \label{self2.eqn},
\eearr
where $c=\pi^2|\tilde v|^2/D^2$, and $b$ determines the modified 
position of the resonance and satisfies $z-\mu-\Delta_R(z)=0$.
The parameters are redefined in units of the bandwidth parameter
$D$ as $\tilde v=\tilde V/D,~v=V/D,~u=U/D,\zeta=\xi/D$.
Now we present our analytic solution of the above self-consistency
equation.

{\em No doping:}
At $\mu=0$ the solution of the equation $z-\mu-\Delta_R(z)=0$ will be given
by $b=0$ and the spectral function takes the simple form 
\be
   \rho_{f}(z)=\frac{1}{\pi}\frac{c|z|}{1+c^2z^4}.
   \label{rhof0.eqn}
\ee
This spectral function despite being drastically
different from a simple Lorentzian form is normalizable,
and is properly normalized to $0.5$ for each spin orientation,
whereas if one uses the $\vk$-independent hybridization
function, instead of the present spectral function at the Dirac
neutrality point, one would get a $\rho_f(z)\propto |z|^{-1}$,
which is not normalizable. The width of the spectrum defined by 
$(\Delta z)^2=\langle z^2\rangle-\langle z\rangle^2$ can be exactly
calculated as 
$
 (\Delta z)^2= \frac{1}{2\pi}\left(\frac{D}{\pi\tilde v}\right)^2
 \ln\left[1+\left(\pi\tilde v\right)^4\right],
$
which for small values of $\tilde v$ can be expanded to give
\be
   T_K =\frac{1}{\sqrt{2\pi}} \pi \tilde v D.
\ee
This striking result must be contrasted with the corresponding result for
metallic hosts, where the width of the resonance is given
by $\pi \tilde v^2 D$~\cite{DSR04}. The width of the Kondo peak
in both metallic and vacant graphene cases relies on a 
non-zero $\tilde v$. For comparable bandwidths and similar values of 
$\tilde v$, the graphene host 
is able to offer a Kondo temperature higher than the metallic host. 
The ratio of the Kondo temperature of a simple metal and a Dirac matter
is given by
\be
   \frac{T_K^{\rm metal}}{T_K^{\rm graphene}}\sim \tilde v.
\ee
This is consistent with the fact that the measured Kondo temperature 
in graphene~\cite{chen} is much higher than Kondo temperatures 
in typical metallic hosts. 
At $\mu=0$ the integral in~\eqref{self2.eqn} can be exactly evaluated to give
\begin{widetext}
\bearr
   \frac{4\zeta^2-u^2}{2uv^2}= \frac{1}{\pi D\tilde v^2}
   \int_{-D}^0  \frac{cz|z|dz}{1+c^2 z^4}=\frac{-\sqrt 2}{8\pi^3\tilde v^3}\times
   \left[\ln\frac{(\pi\tilde v)^2-\sqrt 2\pi\tilde v+1}{(\pi\tilde v)^2+\sqrt 2\pi\tilde v+1} 
   -2\tan^{-1}(1-\sqrt 2\pi\tilde v)+2\tan^{-1}(1+\sqrt 2\pi\tilde v)\right].
   \label{sc.eqn}
\eearr
\end{widetext}
Graphically examining the RHS of the above equation
indicates that it is always negative and has a minimum
at $\tilde v=0$, the Taylor expansion around which gives the
following condition for having a non-zero solution $\tilde v$,
$$
   -\frac{1}{3} < \frac{4\zeta^2-u^2}{2uv^2} < 0 \Rightarrow
   v^2\ge \frac{3}{2}\frac{u^2-4\zeta^2}{u}. 
$$
In the Kondo regime where the single-occupancy of the impurity
site is equivalent to $2\zeta<u$, $u^2-4\zeta^2$ is positive. 
Hence at $\mu=0$, if the hybridization 
strength $v$ is stronger than a minimum $v_{\rm min}$,
\be
   v^2_{\rm min}=\frac{3}{2}\frac{u^2-4\zeta^2}{u}\Rightarrow
   V^2_{\rm min} =-6\eps_d D \left(1+\frac{\eps_d}{U}\right).
\ee
One can have a Kondo resonance at the Dirac point, despite that
the graphene DOS is zero at this point. To estimate the
above minimum, we note that the impurity level $\eps_d$ can be a
very small negative number on the scale of a few meV. This can be simply 
understood as a consequence of a zero-energy mode that has its amplitude on the
majority sublattice pushed to negative-energy side by a repulsive
potential barrier~\cite{SherafatiNJP}. In the case of vacancies
such barrier is expected to be infinitely large, giving rise to negligibly 
small negative value for $\eps_d$. Therefore the condition for the
formation of Kondo resonance at the Dirac neutrality point effectively
becomes $V^2_{\rm min} \approx -6\eps_d D$. For example, assuming 
$-\eps_d\sim 1-10$ meV and $D\sim 10$ eV gives 
$V_{\rm min}\approx 0.25-0.75$eV, which is conceivable in vacant graphene.
This result should be contrasted to the minimal hybridization 
reported in Ref.~\cite{ogata} that requires the hybridization parameter
to be $\sim 7$eV or so.

{\em Doped graphene:}
In this case the integral in the
self-consistency equation~\eqref{self2.eqn} must be approximated.
The modified location of the resonance is approximately
at Fermi level, $b\approx \mu$. Assuming that $\mu$ is small, 
a saddle-point expansion of the integral renders the self-consistency
condition,
$$
   \frac{4\zeta^2-u^2}{4v^2u}=\frac{\pi}{D^3}
   \sum_{m=\pm 1} \frac{z_m^3(z_m-\mu)}{2(z_m-\mu)+6c^2z_m^5}
   \int_{-D}^\mu \frac{sgn(z) dz}{z-z_m},
$$
where two poles dominantly contributing are given by
$z_m\approx \mu+icm\mu^3$ with $m=\pm 1$.
Substituting the above poles and dropping higher
powers of $\mu$ in comparison to lower powers, the self-consistency 
equation simplifies to
$
   \frac{4\zeta^2-u^2}{4v^2u}=
   s\pi \tilde \mu^3\ln[ (\pi\tilde v)^2 \tilde \mu^{2-s} ],
$,
where $s=sgn(\mu)$
and $\tilde\mu=\mu/D$ is the chemical potential in units of $D$.
This equation gives the following explicit solution for the $\tilde v$
characterizing the Kondo temperature for $\tilde \mu\ne 0$, 
\be
   \frac{T_K}{D}\sim\pi\tilde v= |\tilde\mu|^{x_\pm} \exp\left[
   -\frac{1}{2\pi|\tilde\mu|^3}\frac{u^2-4\zeta^2}{4v^2u}\right],
\ee
where $x_\pm=s/2-1$.
The algebraic $|\tilde \mu|^{x_\pm}$
part in our result for $T_K$ resembles that of Ref.~\cite{Bulla} for
the Kondo model. However in the Anderson model we find an extra
exponential dependence extending the result of Ref.~\cite{Bulla}.
In the Kondo regime $u^2>4\zeta^2$, we obtain vanishingly 
small Kondo temperature when the chemical potential approaches zero. 
But the $\tilde\mu=0$ point is singular as we discussed and gives a
Kondo resonance if $V>V_{\rm min}$. 

{\em Conclusion:} Our slave rotor mean field solution for vacancy 
inducing Kondo resonance in graphene explains the possibility of 
a non-zero Kondo temperature at the Dirac point with zero DOS.
We also found a line-shape of the resonance totally different form the usual case,
which is however normalizable and gives rise to a Kondo temperature
in graphene that is linear in effective hybridization $\tilde v$ rather
than the quadratic dependence in ordinary metals. Hence compared with a 
metal with similar bandwidth, a higher $T_K$ can be achieved in graphene.
Upon slight electron or hole doping, the Kondo temperature acquires
exponential dependence on inverse chemical potential with an 
algebraic pre-factor whose power is electron-hole asymmetric. 
In $\tilde \mu=0$ where electron and hole type of excitations
become degenerate, a separate analysis shows that the Kondo
resonance can form under feasible conditions. In the Kondo problem for
alkali metals adsorbed on graphene, despite their hollow site adsorption~\cite{Cohen}
that leads to similar formulation to that of vacancies, their $\eps_d$ values
are around $\sim -1 eV$, which leads to a $V_{\rm min}\sim 3.5$ eV or so 
and hence is not expected to give rise to Kondo resonance at the Dirac point.

S.A.J. thanks the Yukawa Institute for Theoretical Physics for
hospitality within a visiting program.

\end{document}